\documentclass[preprint,10pt]{aastex}
\usepackage{times}

\newcommand{\SMC}{MACHO 98-SMC-1}
\slugcomment{to appear in the Astrophysical Journal (Feb, 2002)}
\shortauthors{GOULD \& AN}
\shorttitle{RESOLVING MICROLENS BLENDS}
\begin{document}

\title{Resolving Microlens Blends Using Image Subtraction}

\author{Andrew Gould and Jin H. An} 
\affil{Department of Astronomy, the Ohio State University,
140 W. 18th Ave., Columbus, OH 43210}
\email{gould,jinhan@astronomy.ohio-state.edu}

\begin{abstract}

Blended light is an important source of degeneracy in the
characterization of microlensing events, particularly in binary-lens
and high magnification events. We show how the techniques of image
subtraction can be applied to form an image of the blend with the source
removed. In many cases, it should be possible to construct images with
very high signal-to-noise ratio.  Analysis of these images can help
distinguish between competing models that have different blend fractions,
and in some cases should allow direct detection of the lens.

\end{abstract}
 
\section{Introduction
\label{sec:intro}}

	Blended light can be a major nuisance in the analysis of microlensing
events, leading to degeneracies in the interpretation of the event parameters.
On the other hand, if unlensed light were detected and could attributed
with good confidence to the lens itself, this would greatly aid in
understanding the physical characteristics of the lens.  The analysis
of blended light is therefore important in several aspects of microlensing.

	In general, microlensing light curves are fit to the functional form,
\begin{equation}
\label{eqn:lc}
F(t) = F_{\rm s} A(t) + F_{\rm b},\qquad A(t)\geq 1
\,,\end{equation}
where $F(t)$ is the flux observed as a function of time,
$F_{\rm s}$ is the flux from the lensed source when it is not magnified,
$F_{\rm b}$ is the flux from any light
that lies within the point spread function (PSF) 
but is not magnified during the event,
and $A(t)$ is the magnification.
The unlensed light $F_{\rm b}$ could come from the lens itself,
from a companion to the lens or to the lensed source
whose projected separation is too great
for the companion to participate in the event,
or from one or more random field stars
that happen to be projected close to the line of sight.  

	When microlensing light curves are fit to models, $F_{\rm b}$ must
be left as an almost completely free parameter.  The only constraint is
that it cannot be negative.  Even this constraint can be violated to a
small extent: microlensing events are found in very crowded fields where
faint stars are typically separated by less than a PSF
and so form part of the ``sky''.  If statistical fluctuations leave
a small hole in this background just at the position of the lensed source, 
then $F_{\rm b}$ can be slightly negative.

	Blending can be degenerate with parameters of the model that
predicts $A(t)$.  For example, in point-source/point-lens events there is a 
continuous degeneracy where higher blending, shorter timescales, and
lower source-lens impact parameters all move in tandem.  This degeneracy
can be resolved with arbitrary precision given arbitrarily good data,
but in realistic cases it places significant limits on the precision
with which the impact parameter and timescale can be determined.  This
in turn limits one's ability to determine the sensitivity of the event
to planetary perturbations \citep{fiveyear-letter,fiveyear-paper}.
For binary-lens events,
there is often a discrete degeneracy between wide-binary and close-binary
solutions, the first example of which \citep{complete-solutions} was discovered
simultaneously with their theoretical prediction \citep{degeneracy}.
These two solutions have substantially different blending parameters
$F_{\rm b}$,
so that if it were possible to rule out (or argue against) one of the
two blending parameters, one could also distinguish between the two solutions.

	There are several tools currently available to deal with blending.
\citet*{DUO2} showed that if a single unlensed star is offset from the 
lensed source by $\theta_{\rm b}$, then it induces an astrometric deviation
on the apparent position of the source, which varies with the magnification:
$\delta\theta = r \theta_{\rm b} (1 - A_{\rm ap}^{-1})$,
where $A_{\rm ap} = (A F_{\rm s} + F_{\rm b})/(F_{\rm s}+F_{\rm b})$
is the apparent magnification, and $r=F_{\rm b}/(F_{\rm s} + F_{\rm b})$
is the unlensed fraction of the baseline flux, $F_{\rm s}+F_{\rm b}$.
This provides a model-independent measure of the parameter combination
$r\theta_{\rm b}$.  The ratio of this quantity measured in two bands provides
a fourth relation among the four flux quantities ($F_{\rm s}$ and $F_{\rm b}$
in each of the two bands), the other three being the two baseline fluxes
and the color of lensed source, which can be determined
from a simple regression between the fluxes in two bands during the event.
Hence, in principle, one can solve for all four quantities.  In practice,
there can be several stars contributing to the unlensed light, 
in which case the 
information extracted by this method would be highly degenerate.  If the
direction of the
astrometric deviation is the same in two bands, this might be taken as
an indication that a single star is dominating the unlensed light, but this
assumption could be violated if there is another blended source whose
position or color is similar to that of the lensed source.
Moreover, crowded-field astrometry is intrinsically
difficult, so that this method may ultimately be limited in the precision
it can reach.

High resolution images by the {\it Hubble Space Telescope} (\emph{HST}) have
been used to resolve blending by random field stars in 8 microlensing events
observed toward the Large Magellanic Cloud
\citep{macho-hst}.
This is a very powerful technique
because even for relatively crowded fields, the chance is small that a random 
star will be both bright enough to affect the event and close enough to
the lensed source to escape detection.  Nevertheless, the method is limited
by the restricted availability of \emph{HST} time.  In particular, 
\emph{HST} observations generally cannot be undertaken until after the 
end of the event when it is too late to influence observational strategy.

	Neither of these methods can distinguish between light from the
lensed source and unlensed light from the lens itself,
nor generally from companions to the lens or the lensed source.
However, future high precision astrometric measurements using
interferometers such as {\it Space Interferometry Mission} (\emph{SIM})
will be able to disentangle these various light sources,
even though they will not be able to resolve them 
\citep*{SIM,astrometric-microlensing,luminous-lens-traj}.
The principle is essentially the same as that of \citet{DUO2},
but the much higher precision allows one to measure the motion of the
centroid of lensed source light, which traces an ellipse.  A luminous lens
will change the parameters of this elliptical motion in a detectable way
\citep{luminous-lens}, and companions to the lens or source will
distort the ellipse into other shapes \citep{binary-source}.
However, it will probably be at least a decade before
this method is used at all, and when it is, it will be feasible only
for a relatively few bright events.

	Here we present a new method to study blends using image subtraction.
There exist many working image subtraction algorithms 
\citep{image-dif-tc,image-dif-al,image-dif,image-dif-wo}.
All share the same 
basic approach.  One first forms a high-quality ``template'' 
from one or several good-seeing images.  Then for each other ``current'' 
image, one convolves the template to the same seeing as the current image,
translates it so the two images are geometrically aligned, and linearly
rescales its flux so that they are photometrically aligned as well.  In
principle, the two images are then identical (up to photon noise) except
where sources have varied.  Hence when the template is subtracted from
the current image, all that remains are a set of (usually) isolated
PSF's at the locations of these variables.  Since the difference image
has the appearance of a high-Galactic-latitude field, photometry is much
easier and more accurate than it is in the original crowded field.

	At first sight, it appears that the unlensed light, and with it the
blend parameter 
$F_{\rm b}$, disappear from the analysis.  In fact, the parameter
$F_{\rm b}$ in equation~(\ref{eqn:lc}) is replaced by an arbitrary offset
in the fit to difference-image photometry, so the degeneracies connected
with the blend persist.  In fact they are somewhat worsened because,
as mentioned above, for crowded-field photometry, one can at least 
constrain $F_{\rm b}$ to be non-negative, 
whereas there are no a priori conditions
upon the corresponding difference-imaging offset parameter.  However,
by doing both types of photometry, and aligning them by linear regression,
one can relate the offset parameter to $F_{\rm b}$, and so recover this
constraint.

	The method we propose is to form a linear combination of
images such that the lensed source is removed from the resulting image,
but all the unlensed sources remain.  The images must be geometrically
and photometrically aligned and convolved to the same seeing before
combining them.  

As we describe in \S~\ref{sec:method},
the resulting image of the unlensed sources
depends explicitly on $A(t)$, that is, on the model of the microlensing
event.   Hence, different models that are consistent with the
photometric data will lead to different images of the unlensed
sources.  As we discuss in \S~\ref{sec:discuss}, some of these
images will be implausible, or even physically impossible.  When
such conflicts exist, they can be used to argue against or rule
out certain classes of models, and 
thus restrict the allowed space of solutions.
Some required statistics results are derived in an Appendix.

\section{Constructing an Image of the Unlensed Sources
\label{sec:method}}

Consider a series of $n$ images \{$I_i(x,y)$, $i=1,\ldots,n$\},
as functions of pixel position $(x,y)$.  For simplicity,
we will assume that these images have already been convolved to the
same seeing, geometrically aligned, and linearly rescaled, so that they
may be directly compared with one another.  All the sources in the image
will be assumed to be constant, except the microlensed source which
varies with magnification $A_i=A(t_i)$.  We adopt $A(t)$ from the 
model of the light curve under consideration.  That is, different models
lead to different images of the unlensed sources, only one of which
can be correct.  We show in \S~\ref{sec:discuss} how to use the
resulting images to distinguish among competing models.

Let $B(x,y)$ be an arbitrary linear combination of images
\begin{equation}
\label{eqn:bofxy}
B(x,y) = \sum_{i=1}^n a_i(x,y) I_i(x,y)
\,,\end{equation}
where the $a_i(x,y)$ are coefficients that vary both as a function of
pixel position in the image $(x,y)$, and as a function of image number $i$,
but subject to the two constraints
\begin{equation}
\label{eqn:constraints}
\sum_{i=1}^n a_i(x,y) = 1
\,;\ \ \
\sum_{i=1}^n a_i(x,y) A_i = 0
\ .\end{equation}
Then $B$ will be an image of the field with the lensed source removed.  
That is,
the first constraint insures that all constant sources will be retained
in $B$, while the second insures that the microlensed source will be
deleted.  With the microlensed source removed from $B$, the neighboring
unlensed sources can be studied more closely.

For the simplest case $n=2$, the constraints 
(eqs.~[\ref{eqn:constraints}]) then completely determine the $a_i$, 
\begin{equation}
\label{eqn:aoneatwo}
a_1(x,y) = \frac{A_2}{A_2 - A_1}
\,;\ \ \
a_2(x,y) = \frac{-A_1}{A_2 - A_1}
\,,\end{equation}
which are then independent of position.  This implies
\begin{equation}
\label{eqn:nequalstwo}
B = \frac{A_2 I_1 - A_1 I_2}{A_2 - A_1}
\ \ \ \ {\rm for}\ n=2
\ .\end{equation}

\begin{figure}
\plotone{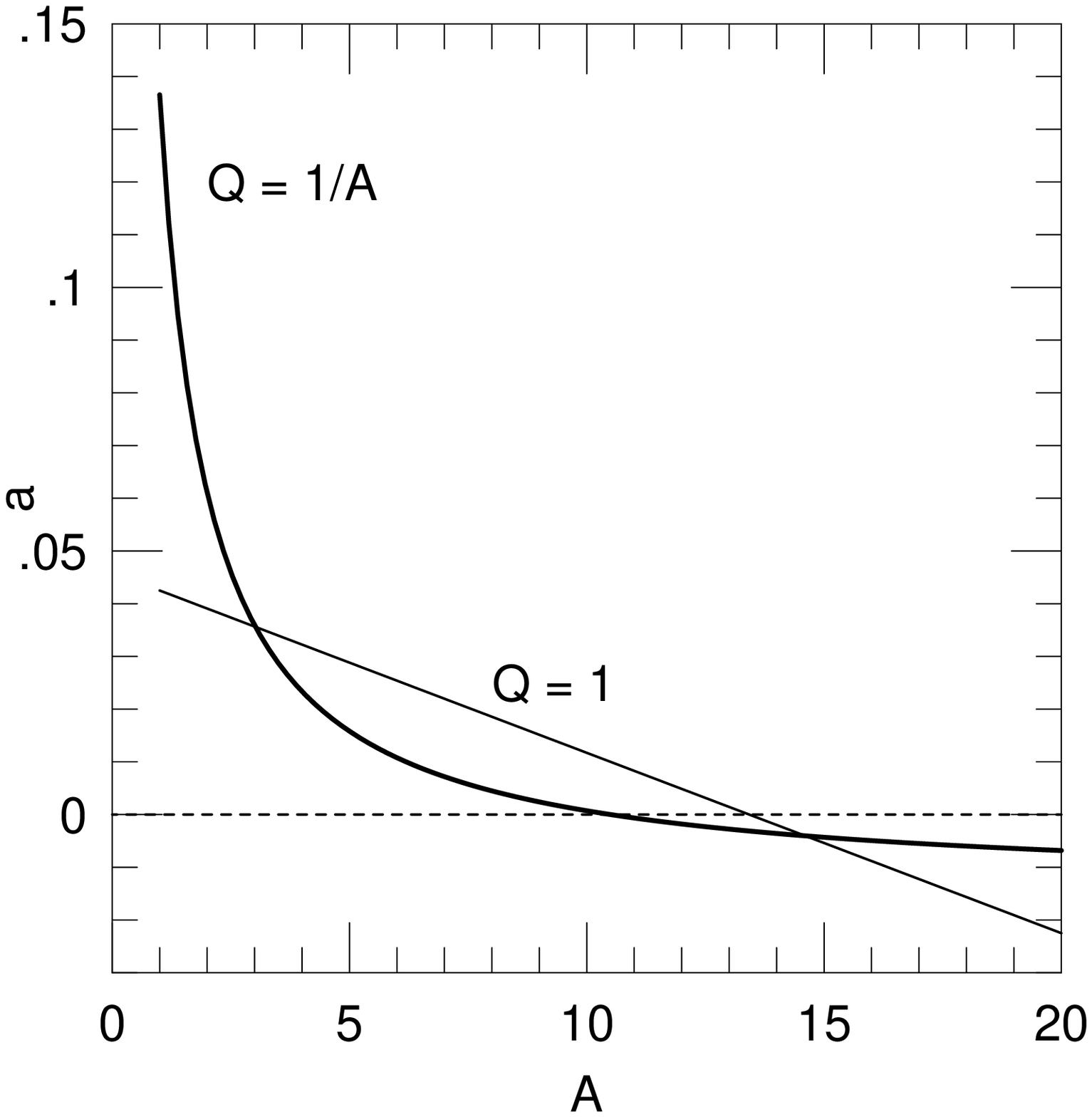}
\caption{\label{fig:one}
Optimal weighting factors, $a$, for combining images to produce a source-free
image of the unlensed light in a microlensing event, as functions of the
magnification, $A$, of each image.  The examples shown are for a total
of $n=100$ images with magnifications uniformly distributed over the interval
$1\leq A \leq 20$.  The solid line shows the factors for the case
where the flux
errors are independent of the magnification (constant $Q$), which is
appropriate when even the highly magnified source is below the sky.
The bold curve shows the factors for the case where
the flux errors are proportional
to the square root of the magnification
($Q\propto 1/A$), which is appropriate when the
source itself is above the sky.
}\end{figure}

Forming $B$ as a combination of only two images may well suffice
for many applications.  In this case, the images should be chosen to
both have very good seeing and very different magnifications.
However, there may be other cases where one wants to obtain a higher 
signal-to-noise ratio (S/N) for $B$ by combining a large number of images.
Let $\sigma_i(x,y)$ be the error in $I_i(x,y)$.  For example, $\sigma_i(x,y)$
might be given by the photon noise in the $(x,y)$ pixel of image $I_i$,
or it might contain additional sources of error due to other causes.
And define,
\begin{equation}
\label{eqn:qidef}
Q_i(x,y) \equiv \frac{1}{[\sigma_i(x,y)]^2}
\ .\end{equation}
Then the error in $B(x,y)$ is given by,
\begin{equation}
\label{eqn:sigmabxy}
\sigma[B(x,y)] = \sqrt{\sum_{i=1}^n \frac{[a_i(x,y)]^2}{Q_i}}
\ .\end{equation}
Since the form of $B$ is fixed, maximizing the S/N in $B$ is
equivalent to minimizing equation~(\ref{eqn:sigmabxy}) subject to the
constraints (eqs.~[\ref{eqn:constraints}]).  It is straight forward to show
(see Appendix) that this is accomplished when
\begin{equation}
\label{eqn:aofi}
a_i(x,y) = 
\frac
{\langle A^2 Q(x,y)\rangle Q_i(x,y) - \langle A Q(x,y)\rangle A_i Q_i(x,y)}
{n[\langle A^2 Q(x,y)\rangle \langle Q(x,y)\rangle
-\langle A Q(x,y)\rangle^2]}
\,,\end{equation}
where
\begin{equation}
\label{eqn:vevg}
\langle G(x,y)\rangle \equiv \frac{1}{n}\sum_{i=1}^n G_i(x,y)
\ .\end{equation}
It is not immediately obvious, but equations~(\ref{eqn:aofi}) reduces
to equation~(\ref{eqn:aoneatwo}) for the special case of $n=2$.
By direct substitution of equation~(\ref{eqn:aofi}) 
into equation~(\ref{eqn:sigmabxy}),
\begin{equation}
\label{eqn:sigmabeval}
\sigma[B(x,y)] = 
\frac{1}{\sqrt{n}}
\left[\langle Q(x,y)\rangle -
\frac{\langle A Q(x,y)\rangle^2}{\langle A^2 Q(x,y)\rangle}
\right]^{-1/2}
\ .\end{equation}
For the special case of $n=2$, this becomes
\begin{equation}
\label{eqn:sigmabonetwo}
\sigma[B(x,y)] = 
\frac{[A_2^2/Q_1(x,y) + A_1^2/Q_2(x,y)]^{1/2}}{|A_2-A_1|}
\,,\end{equation}
which confirms that for the case $n=2$ it is best to choose images
with the widest difference in magnifications.  Indeed, even for the
case $n>2$, equation~(\ref{eqn:aofi}) shows that images at the
extreme ranges of magnification are automatically given more weight
than those with intermediate magnifications.  To illustrate this
concretely, we show in Figure~\ref{fig:one} 
two cases, $Q\propto A^{-1}$ and constant $Q$,
which are respectively appropriate for the limits where the magnified
source is always above and always below the sky.  
We choose $n=100$ and assume that the magnifications 
$A_i$ are uniformly distributed between 1 and 20.

Of course, one could choose the coefficients $a_i$ to be independent
of position.  In this case, one would apply equation~(\ref{eqn:aofi})
by adopting a set of $Q_i$ that are representative of the
images, for example the $Q_i$ averaged over the PSF of the magnified
source, or of the central pixel.  However, there does not appear 
to be any compelling reason to do this.  Moreover, the strong difference
in the functional forms shown in Figure~\ref{fig:one} implies that the
optimal linear combinations may be very different near the 
center of the lensed source (which tends toward being above the sky)
than they are near the wings (which tends toward being below the sky).

We note that an image of the (unmagnified) source, $S$, can also be constructed
in a similar manner. Such an image may be useful when studying the
image of the unlensed light. With alternative constraints
\begin{equation}
\sum_{i=1}^n a^\prime_i(x,y) = 0
\,;\ \ \
\sum_{i=1}^n a^\prime_i(x,y) A_i = 1
\,,\end{equation}
one obtains in place of equations~(\ref{eqn:nequalstwo}),
(\ref{eqn:aofi}), (\ref{eqn:sigmabeval})
and (\ref{eqn:sigmabonetwo});
\begin{mathletters}
\label{eqn:sourceimage}
\begin{equation}
S = \frac{I_2 - I_1}{A_2 - A_1}
\ \ \ \ {\rm for}\ n=2
\,,\end{equation}
\begin{equation}
a^\prime_i(x,y) =
\frac
{\langle Q(x,y)\rangle A_i Q_i(x,y) - \langle A Q(x,y)\rangle Q_i(x,y)}
{n[\langle A^2 Q(x,y)\rangle \langle Q(x,y)\rangle
-\langle A Q(x,y)\rangle^2]}
\,,\end{equation}
\end{mathletters}
\begin{mathletters}
\begin{equation}
\sigma[S(x,y)] =
\frac{1}{\sqrt{n}}
\left[\langle A^2 Q(x,y)\rangle -
\frac{\langle A Q(x,y)\rangle^2}{\langle Q(x,y)\rangle}
\right]^{-1/2}
\,,\end{equation}
\begin{equation}
\sigma[S(x,y)] =
\frac{[1/Q_1(x,y) + 1/Q_2(x,y)]^{1/2}}{|A_2-A_1|}
\ \ \ \ {\rm for}\ n=2
\ .\end{equation}
\end{mathletters}

\section{Discussion
\label{sec:discuss}}

	The method outlined in \S~\ref{sec:method} is likely to
prove useful primarily in cases where
the uncertainty in the photometric determination of blending is large,
either because of intrinsic degeneracies or because the baseline is very
faint.  The first condition applies mainly to caustic-crossing
binary lenses, while the second applies to high-magnification events.  
Caustic-crossing binaries are themselves a major source of high-magnification
events, so binaries are the singled out for both reasons.  The spectacular
binary event \SMC, which had an extremely faint ($I\sim 22$)
source and maximum magnification of $A\sim 100$, is an excellent example:
five microlensing groups combined their precise and extensive
data sets, but were still not able to distinguish between two solutions
with values of $F_{\rm b}$ that differed by a factor of two
(\citealt{combined-analysis} and references therein).

	However, the precise determination of the blending with
the aid of image subtraction can also significantly
impact generic high-magnification events, which play a major
role in planet searches \citep{fiveyear-letter,fiveyear-paper}.
Such events are exceptionally sensitive to 
planetary companions of the primary lens,
because both the planetary caustic \citep{planetary-caustic} 
and the central caustic \citep{central-caustic}
are much more likely to cause planetary perturbations
than in typical events.  For this reason, these events are often monitored 
more intensively than typical events, which further increases the
sensitivity of these events to planet detection.
No planets have yet been discovered,
but accurate estimates of the blending are essential to place
upper limits on the presence of planets in the cases of non-detections.
For example, \citet{fiveyear-paper} excluded from their analysis all
events where the uncertainty in the lens-source impact parameter was
greater than 50\% because these uncertainties led to a significant
(and difficult to determine) underestimation in the events' sensitivity
to planets.  A 50\% uncertainty in impact parameter corresponds to the
difference between a model with $F_{\rm b}=0$ and one with 
$F_{\rm b}=F_{\rm s}$.  (About one quarter of all the events included
in their analysis had impact-parameter uncertainties greater than 25\%.)
Hence, even though the degeneracy is continuous
in this case (as opposed to the discrete degeneracy discussed above for
binaries), the uncertainty in the blending parameter $F_{\rm b}$
derived solely from
the analysis of the light curve, can still be quite large.  This implies
that additional information about the blend acquired from the analysis
of the images can be important in reducing or resolving the degeneracy.

	How can such information be extracted?  Primarily by investigating
whether the unlensed light 
(or a significant component of it) has both the same color
and the same position as the source.  Recall that the source color can
be determined very precisely by regression of the flux in two bands, while
the source position can be determined very precisely using image subtraction.

	More specifically, if the lens model has incorrectly
estimated the flux of the lensed source, 
then it will attribute the difference between
the true and estimated $F_{\rm s}$ to the blend parameter $F_{\rm b}$.
Thus, for instance,
there may be no actual unlensed light, but an incorrect fit
will yield a finite $F_{\rm b}$.  If $F_{\rm b}$ is significantly negative, 
the error of the model will be manifest.  But if $F_{\rm b}$ is positive,
then no such argument can be made.  However, in this case one can still 
perform another test: an unlensed source
that was inferred on the basis of an incorrect
model would have exactly the same color as the lensed source.  Hence, identical
colors would be a strong hint of a wrong model.  Unfortunately, this
test is not definitive.  On the one hand, an unlensed star could have
the same color as the lensed 
source, so that a common color would not prove that
the inferred unlensed light was an artifact.  On the other hand,
even if the color is found to be different,
this would prove only that there was \emph{some} real source of unlensed light
but it would not prove that the model $F_{\rm b}$ was correct.  That is, a 
significant (either positive or negative) fraction of the flux attributed by
the model to unlensed light could still belong to the lensed source.  

	If the derived $F_{\rm b}$ 
is purely an artifact of the model, then the procedure
outlined in \S~\ref{sec:method} will yield an isolated ``unlensed source'' with
exactly the same position as the lensed source.  The combination of this
and the fact that it had the same color as the lensed 
source would be convincing
proof that it was an artifact.  On the other hand, a real unlensed source 
that was
displaced from the lensed source by as little as $\sim 10\%$ of the PSF would
easily be distinguishable from an artifact.  For example, if this unlensed
source was subtracted from a scaled image of the lensed source 
(see eqs.~[\ref{eqn:sourceimage}]),
the result would be a ``shadowed mountain'' effect
which is familiar from subtraction of misaligned images.

	If the image of the unlensed light
consisted of an isolated point source
at the same position as the lensed source but having a different color
from it, then one could conclude that this was probably the lens itself
(or possibly a companion to the lens or source), since the chance of
a random field star being aligned with the source to within
$\la 10\%$ of the seeing disk is small.

	Of course, it is possible that there is more than one unlensed 
source, and if these are too closely packed together, it will be difficult to 
make sense of them, even with the lensed source removed.  However, regardless
of the configuration of the unlensed sources, 
it will be easier to disentangle them
in the image with the lensed
source removed, than it is in an image containing the lensed
source.  This will be especially true when all the sources are very 
faint.  In this case, the high S/N image of the unlensed sources
formed by combining 
many individual images will mark an exceptional improvement.

	It is important to keep in mind that all the images that are
combined to form an image of the unlensed light 
must be convolved to the \emph{worst}
seeing of the lot.  Hence, there is an inevitable tradeoff in forming
a high S/N image between increasing the S/N and decreasing the resolution.
One should therefore rank order the images by seeing, and set the
threshold at various values to determine which ensemble of images produces
the best image of the unlensed light.

\acknowledgements
{\center \bf ACKNOWLEDGEMENTS}

This work was supported in part by NSF grant AST~97-27520 and in part by
JPL contract 1226901.

\clearpage
\newpage
\appendix
{\center \bf APPENDIX}

\section{Minimization with Constraints}

We use Lagrange multipliers to evaluate
the $n$-dimensional vector $\{\tilde a_i\}$
that minimizes the quadratic function
${\mathcal{H}} (\{a_i\})=\sum_{i,j=1}^n b_{ij} (a_i-a^0_i) (a_j-a^0_j) 
+ {\mathcal{H}}_0$,
subject to the $m$ constraints
$\sum_{i=1}^n a_i\alpha_i^k = z^k$ ($k=1,\ldots,m$).
Here the $a^0_i$, the $b_{ij}$, and ${\mathcal{H}}_0$ are constants.
At this minimum, $\nabla {\mathcal{H}}$ must lie in the $m$-dimensional
subspace spanned by the constraint vectors $\{\alpha^l_i\}$, i.e.,
$\sum_{j=1}^n b_{ij}(\tilde a_j-a^0_j) + \sum_{l=1}^m D^l \alpha^l_i = 0$,
or
\begin{equation}
\label{eqn:aisolve}
\tilde a_i = a^0_i- \sum_{l=1}^m D^l \kappa^l_i
\,;\ \ \
\kappa^l_i \equiv \sum_{j=1}^n c_{ij} \alpha^l_j
\,,\end{equation}
where $(c_{ij})\equiv (b_{ij})^{-1}$,
and the $D^l$ are constants to be determined.  
Multiplying equation~(\ref{eqn:aisolve}) by each of the $\alpha^k_i$ yields
a set of $m$ equations,
\begin{equation}
\label{eqn:calb}
\sum_{l=1}^m C^{kl} D^l = \sum_{i=1}^n a^0_i \alpha^k_i - z^k
\,;\ \ \
C^{kl} \equiv \sum_{i=1}^n \alpha^k_i \kappa^l_i 
= \sum_{i,j=1}^n c_{ij} \alpha^k_i \alpha^l_j
\,,\end{equation}
which can be inverted to solve for the $D^k$,
\begin{equation}
\label{eqn:cala}
D^k = \sum_{l=1}^m B^{kl} \left( \sum_{i=1}^n a^0_i \alpha^l_i - z^l \right)
\,;\ \ \
\left(B^{kl}\right) = \left(C^{kl}\right)^{-1}
\ .\end{equation}
To obtain equation~(\ref{eqn:aofi}) in \S~\ref{sec:method},
one sets ${\mathcal{H}}_0=0$, $a_i^0=0$, $b_{ij}=\delta_{ij}/Q_i$,
$m=2$, $\alpha^1_i = 1$, $\alpha^2_i=A_i$, $z^1=1$, $z^2=0$,
and then substitutes into equations~(\ref{eqn:aisolve})
and (\ref{eqn:cala}).

Note that for the special case where ${\mathcal{H}}$ can be interpreted as
$\chi^2$ (not of direct concern here, but of general interest), 
$c_{ij}$ is the covariance matrix of the unconstrained parameters $a_i$,
i.e., $c_{ij}={\rm cov}(a_i,a_j)
=\langle a_i a_j \rangle - \langle a_i \rangle \langle a_j \rangle$.
One then finds by direct substitution that
${\rm cov}(D^k,a^0_i)=\sum_{l=1}^m B^{kl}\kappa_i^l$, so that
the covariances of the constrained parameters,
$\tilde c_{ij} = {\rm cov}(\tilde a_i,\tilde a_j)$, are given by
\begin{equation}
\label{eqn:cijsolve}
\tilde c_{ij} = c_{ij} -\sum_{k,l=1}^m B^{kl} \kappa_i^k \kappa_j^l
\ .\end{equation}

To further specialize to an important subcase, let $\alpha^k_i = \delta_{ik}$.
Then $\kappa^k_i = c_{ik}$ and $C^{kl}=c_{kl}$, hence $B^{kl}= \hat b_{kl}$,
where $(\hat b_{kl}) = (\hat c_{kl})^{-1}$ and 
$\hat c_{kl}$ is the (unconstrained) $m\times m$ covariance matrix
restricted to the $m$ parameters that are to be constrained.
Thus, for this special case, equations~(\ref{eqn:aisolve}), 
(\ref{eqn:cala}), and (\ref{eqn:cijsolve}) become
\begin{equation}
\label{eqn:simplified}
\tilde a_i = a_i^0 - \sum_{k,l=1}^m c_{ik} \hat b_{kl} (a^0_l - z^l)
\,;\ \ \
\tilde c_{ij} = c_{ij} - \sum_{k,l=1}^m \hat b_{kl} c_{ik} c_{jl}
\ .\end{equation}

\clearpage
\newpage

\end{document}